\documentclass[reprint,aip, apl]{revtex4-1}
\usepackage{graphicx}
\usepackage{bm}
\usepackage[usenames,dvipsnames]{color}
\usepackage{soul}
\usepackage{color}

\begin{document}

\title{Charge Sensing in Intrinsic Silicon Quantum Dots}
\author{G. J. Podd}
\email{gp278@cam.ac.uk}
\affiliation{Hitachi Cambridge Laboratory, Cavendish Laboratory, Cambridge, U.K.}
\author{S. J. Angus}
\affiliation{School of Physics, The University of Melbourne, Melbourne, Australia}
\author{D. A. Williams}
\affiliation{Hitachi Cambridge Laboratory, Cavendish Laboratory, Cambridge, U.K.}
\author{A. J. Ferguson}
\affiliation{%
Microelectronics Research Centre, Cavendish Laboratory, Cambridge, U.K.}%
\date{\today}

\begin{abstract}

We report charge sensing measurements on a silicon quantum dot with a nearby silicon single electron transistor (SET) acting as an electrometer.  The devices are electrostatically formed in bulk silicon using surface gates. We show that as an additional electron is added onto the quantum dot, a charge is induced on the SET of approximately 0.2$\it{e}$. These measurements are performed in the many electron regime, where we can count in excess of 20 charge additions onto the quantum dot.

\end{abstract}

\maketitle

Silicon is a promising material in which to build a quantum computer \cite{kane_silicon-based_1998}. This is primarily due to the long coherence time for electron spins, as demonstrated by electron spin resonance measurements on dopant ensembles \cite{feher_electron_1959,tyryshkin_electron_2003},  and also expected for electrons in quantum dots (QDs) \cite{Tyryshkin2006}. In recent experiments, the hyperfine interaction was demonstrated to limit spin coherence for electrons confined in gallium arsenide QDs \cite{petta_coherent_2005}. With the motivation of going to nuclear spin free materials, much research effort is currently focussed on accessing the spin of electrons confined in silicon-based nanostructures \cite{liu_pauli-spin-blockade_2008,shaji_spin_2008}.

Our approach to confining single electrons uses electrostatic gates to define QDs in intrinsic silicon \cite{angus_gate-defined_2007}. Related techniques are also possible in silicon-on-insulator (SOI) \cite{fujiwara_single_2006} and silicon-germanium \cite{klein_coulomb_2004}.  For both QDs and dopants a sensitive charge detector is an important experimental tool \cite{field_measurements_1993}. It allows confirmation that a single electron is confined in the potential. Such confirmation is also possible by electrical transport through a dot, but the geometry must ensure that there is still a measureable current at low electron numbers. An equally important advantage of the charge sensor is that it enables single shot measurement of electron spin states \cite{elzerman_single-shot_2004} - a useful property for quantum computation. In silicon-germanium a quantum point contact has been used for charge sensing on QDs in the few electron regime \cite{simmons_single-electron_2007}. In SOI, electron occupancy on an isolated node has been measured with a single electron transistor (SET) at room temperature \cite{nishiguchi_multilevel_2004}, but not probed at cryogenic temperatures. In a narrow silicon metal-oxide-semiconductor field-effect transistor device, an aluminium SET has been used to detect charges on a self-aligned silicon SET  \cite{sun_coulomb_2007}. In this Letter, we demonstrate charge sensing of a QD by a nearby SET co-fabricated in intrinsic silicon.

\begin{figure}
	\centering
		\includegraphics[width=7.5cm]{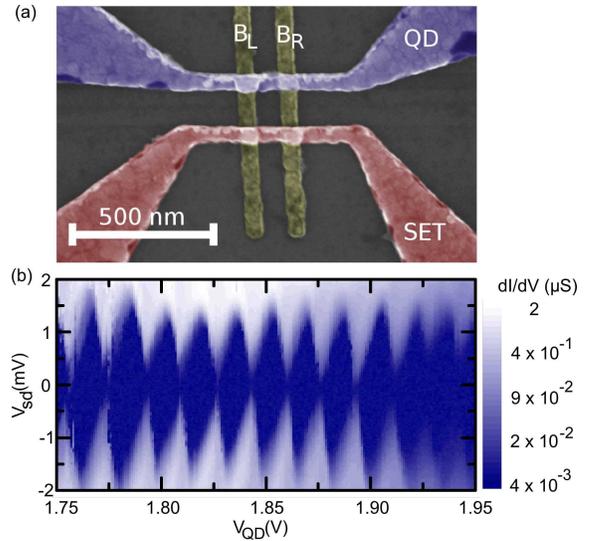}
	\caption{(Color online) (a) False colour scanning electron microscopy (SEM) image of a representative device showing both SET and QD. The top gate of the SET is shown in red, with the top gate of the quantum dot device in blue.  The two yellow barrier gates are common to both devices and are used to locally deplete the 2DEG. (b) Electrical transport measurement of the quantum dot device taken with barrier potentials $B_L = B_R = 0.75V$ over a top gate range representative of that used for charge sensing. The SET gate potential is held at 0~V for this measurement.}
	\label{fig:fig1}
\end{figure}

Previously we have measured similar devices via electrical transport and observed both Coulomb blockade and orbital excited states in the island \cite{angus_gate-defined_2007}. In addition we have embedded these structures in a radio-frequency circuit \cite{angus_silicon_2008} and performed high-bandwidth charge measurement with the device configured as a radio-frequency SET \cite{schoelkopf_radio-frequency_1998}. In this work we operate the QD and SET at low frequencies using both dc and lock-in amplifiers. The measurements presented here were performed at T~$\leq$~100mK in a 0.2~T magnetic field, which is used to suppress superconductivity in the AlSi bondpads.

Figure 1(a) shows a scanning electron microscope (SEM) image of a typical device. The fabrication process has been described previously \cite{angus_silicon_2008}. A two-dimensional electron gas (2DEG) is electrostatically induced using the top gates, whilst the two barrier gates, (B$_L$ and B$_R$ in Fig. 1(a)) are used to locally deplete the 2DEG to form tunnel barriers. The patterned linewidth of the gates is~50nm, and the barrier gates are spaced~85nm apart, resulting in QD dimensions of approximately 50 x 85nm.  The distance between the dot and electrometer is 125nm.

\begin{figure}
	\centering
		\includegraphics[width=7.5cm]{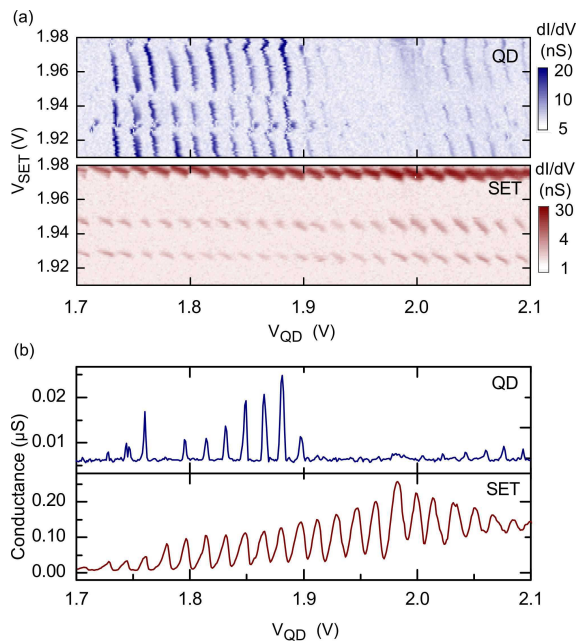}
	\caption{(Color online) Electrical transport measurements of both the SET and QD as a function of QD gate and SET gate.  (a) Conductance of the QD (Top) and SET (Bottom) measured simultaneously. A lock-in excitation of 300$\mu$V was used on the QD and 100$\mu$V on the SET. (b) Line plots extracted from (a) at $V_{SET} = 1.975V$ for both the quantum dot (top) and SET (bottom).}
	\label{fig:fig2}
\end{figure}

For the measurements described here, the potential on each barrier gate was set to 0.75~V.  This value was chosen as a compromise allowing multiple Coulomb blockade oscillations to be seen in the QD device, whilst still being able to observe a number of oscillations in the SET.  Such a compromise between the performance of the QD and SET devices was necessary due to the barriers being common to both QD and SET, which has the advantage of ease of fabrication but places constraints on the device operation.

With the barrier gate potential fixed, the SET and QD were then characterised independently of one another. This is achieved by setting the top gate potential of the device not being measured to 0~V.  Fig. 1(c) shows Coulomb blockade diamonds taken at a lock-in excitation voltage of 200 $\mu$V for the QD and reveals a charging energy of $e^{2}/C_{\Sigma}$ =  1.4~meV, and a gate period of 17~mV.  This corresponds to a gate capacitance of 9.6~aF, in excellent agreement with the value of 9.8~aF calculated using a simple parallel plate capacitor model ($C = \epsilon_{r}\epsilon_{0}A/d$ where $A$ is the area of the dot, $d$ the thickness of the oxide and $\epsilon_{r} = 3.9$).  Similarly, Coulomb blockade diamonds taken for the SET (not shown) give a value of the charging energy of 1.35~meV with a gate period of 20~mV corresponding to a value of the capacitance of the top gate to the SET of 8~aF.

Charge sensing of the QD by the SET  was then performed by measuring both devices simultaneously, shown in Figure 2.  For these measurements the top gate of the QD device (V$_{QD}$) was swept over a range corresponding to more than 20 charge additions to the QD, whilst the SET gate (V$_{SET}$) was incremented over a range of three Coulomb blockade oscillations. In this device we were not able to operate the QD in the few-electron regime due to the constraints imposed by the mutual barriers. We estimate, using the Coulomb blockade peak spacing and the threshold voltage of the device, that we have of the order 25 electrons present in the dot at the lowest gate potential used here. The near vertical (horizontal) lines in the top (bottom) plot of Figure 2(a) are the Coulomb blockade oscillations in the QD (SET). As an electron is added to the QD there is a shift in the SET peak position.  This shift corresponds to approximately 0.2\textit{e}.  The mutual action of the SET on the QD is seen by the shifts in the vertical lines, which are more clearly evident in Figure 3.

Due to variation in the coupling of the QD to the leads, a modulation of the peak height is observed for the QD conductance over the gate range shown in Fig. 2(b). This variation in peak amplitude for the QD gives an excellent demonstration of the benefits of an electrometer as a probe of the charge state of the QD. In the range $1.91~V \leq V_{QD} \leq 2~V$ the current through the QD approaches the noise level of the measurement, and Coulomb blockade peaks are difficult to resolve.  The simultaneous measurement of the current through the SET charge sensor reveals the continued periodic addition of electrons onto the QD in this region with the consistently large charge sensing signal.

\begin{figure}
	\centering
		\includegraphics[width=7.5cm]{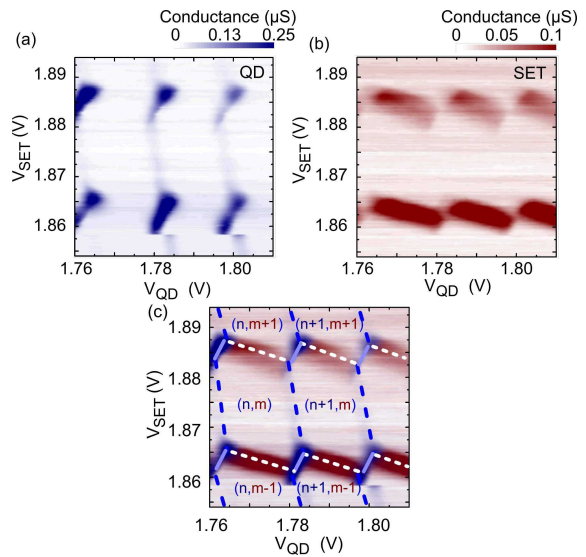}
	\caption{(Color online) Direct conductance measurements of the (a) QD and (b) SET over a small gate range.  (c) Schematic of the 'honeycomb' pattern for coupled quantum dots overlaid on a superposition of plots (a) and (b).}
	\label{fig:fig3}
\end{figure}

The capacitive coupling between the two dots results in the typical honeycomb stability plot of a double quantum dot system, as shown in Fig 3(c). The coupling capacitance between the SET and the QD can be determined from this stability plot and is found to be 20~aF.

An interesting feature of the data is the conductance in the vicinity of the triple points.  The expectation was for a finite but suppressed current in each device between the triple points due to a `turnstile' effect whereby an electron tunnelling off one dot must coincide with an electron tunnelling onto the other in order to observe a current \cite{chan_strongly_2002}. Figure 2(a) shows that we observe this behaviour at all gate potentials in the SET, however in the less conductive QD there are certain gate configurations, for example those shown in Fig. 3,  where rather than a suppression, an enhancement of up to an order of magnitude in the conductance is seen around the triple points.  Here we examine possible mechanisms that could result in the observed enhancement.

Firstly we consider heating effects due to the current through the SET \cite{Krupenin1999}.  In the configuration described here, the current carrying channels of each device are formed in the same silicon crystal just 125~nm apart, so we expect thermal effects to be present, leading to an enhancement of the current.

Another possibility is that rather than simply capacitive coupling between devices, there could be some tunnel coupling due to their proximity.  This would require particular arrangements of barrier transparencies, but under the correct conditions there could be a current path from the SET to QD resulting in an increased current in the QD drain. We have also considered enhanced cotunneling mechanisms as in systems of parallel dots in the Kondo regime enhanced current near the triple points has been reported \cite{Holleitner2002}.

The final mechanism that we examine is that of charge pumping.  In this case electrons tunneling on and off the SET, which is strongly coupled to its leads, act as a time-varying gate potential on the QD.  The device could then be operating in a similar manner to a single parameter charge pump or ratchet \cite{kaestner2008, fujiwara2008}, manifesting in a driven current through the QD whenever the electron number on the SET is changing.

Each of these mechanisms could contribute in some way to the enhancement in the QD.  The configuration of barrier transparencies required for the tunnel coupling would make that unlikely and  heating alone would seem insufficient to explain the large enhancement seen.  However with this device geometry it is difficult to identify the dominant cause. Replacing the barrier gates that are shared between QD and SET  with independent gates will give a more tunable device and allow these effects to be investigated further.

In conclusion, we have demonstrated charge sensing of a silicon quantum dot by an integrated silicon electrometer.  The device has shown a large (0.2\textit{e}) charge sensing signal over a large number of oscillations. The limitations imposed by common barrier gates mean that this device was used to prove the principle of operation, and in future we will implement separate barrier gates for QD and SET with the intention of studying the few-electron regime, of interest for quantum information processing.

This work was partly supported by Special Coordination Funds for Promoting Science and Technology (Japan). S. J. Angus acknowledges financial support from the University of Melbourne and the Australian Research Council.

\bibliographystyle{apsrev}

\end{document}